\documentclass[11pt]{article}
\usepackage{moriond,epsfig}
\usepackage{epsf,amsmath}

\newcommand{\up}{\uparrow}
\newcommand{\down}{\downarrow}

\def\Journal#1#2#3#4{{#1} {\bf #2}, #3 (#4)}

\def\EPJC{{\em Eur. Phys. J.} C}

\def\NIMA{{\em Nucl. Instrum. Methods} A}
\def\NPB{{\em Nucl. Phys.} B}
\def\PLB{{\em Phys. Lett.}  B}
\def\PRL{\em Phys. Rev. Lett.}
\def\PRD{{\em Phys. Rev.} D}
\def\ZPC{{\em Z. Phys.} C}

\def\be{\begin{equation}}
\def\ee{\end{equation}}
\def\bea{\begin{eqnarray}}
\def\eea{\end{eqnarray}}

\begin{document}
\vspace*{4cm}
\title{TRANSVERSE SPIN PHYSICS AT HERMES}

\author{ \underline{U.~ELSCHENBROICH}$^\mathrm{a}$,
  G.~SCHNELL$^\mathrm{b}$, R.~SEIDL$^\mathrm{c}$ \\
  (on behalf of the HERMES--Collaboration)}

\address{\vspace*{\baselineskip}
  $^\mathrm{a}$Vakgroep Subatomaire en Stralingsfysica,\\
  Universiteit Gent, Belgium\\
  $^\mathrm{b}$Department of Physics,\\
  Tokyo Institut of Technology, Japan\\
  $^\mathrm{c}$Physikalisches Institut II, \\
  Friedrich--Alexander--Universit\"at 
  Erlangen--N\"urnberg, Germany}

\maketitle\abstracts{Single--spin asymmetries in semi--inclusive pion 
  production are measured by the HERMES experiment for the first time, with 
  a transversely polarised hydrogen target. Two different sine--dependencies 
  are extracted which can be related to the quark distributions 
  \textit{transversity} $h_1(x)$ and the \textit{Sivers} function 
  $f_{1T}^\perp(x)$.}

\section{Introduction}

Deep inelastic scattering (DIS) as a probe to investigate the structure of the 
nucleon has provided exciting, detailed results in the last decades.
This process, in which a lepton scatters off a nucleon via the exchange of a 
single virtual photon, remains a successful tool to gain novel 
information about the inside of the nucleon.

\begin{figure}	
  \unitlength1cm
  \begin{picture}(17,4.)
    \put(1.,0.5){\psfig{figure=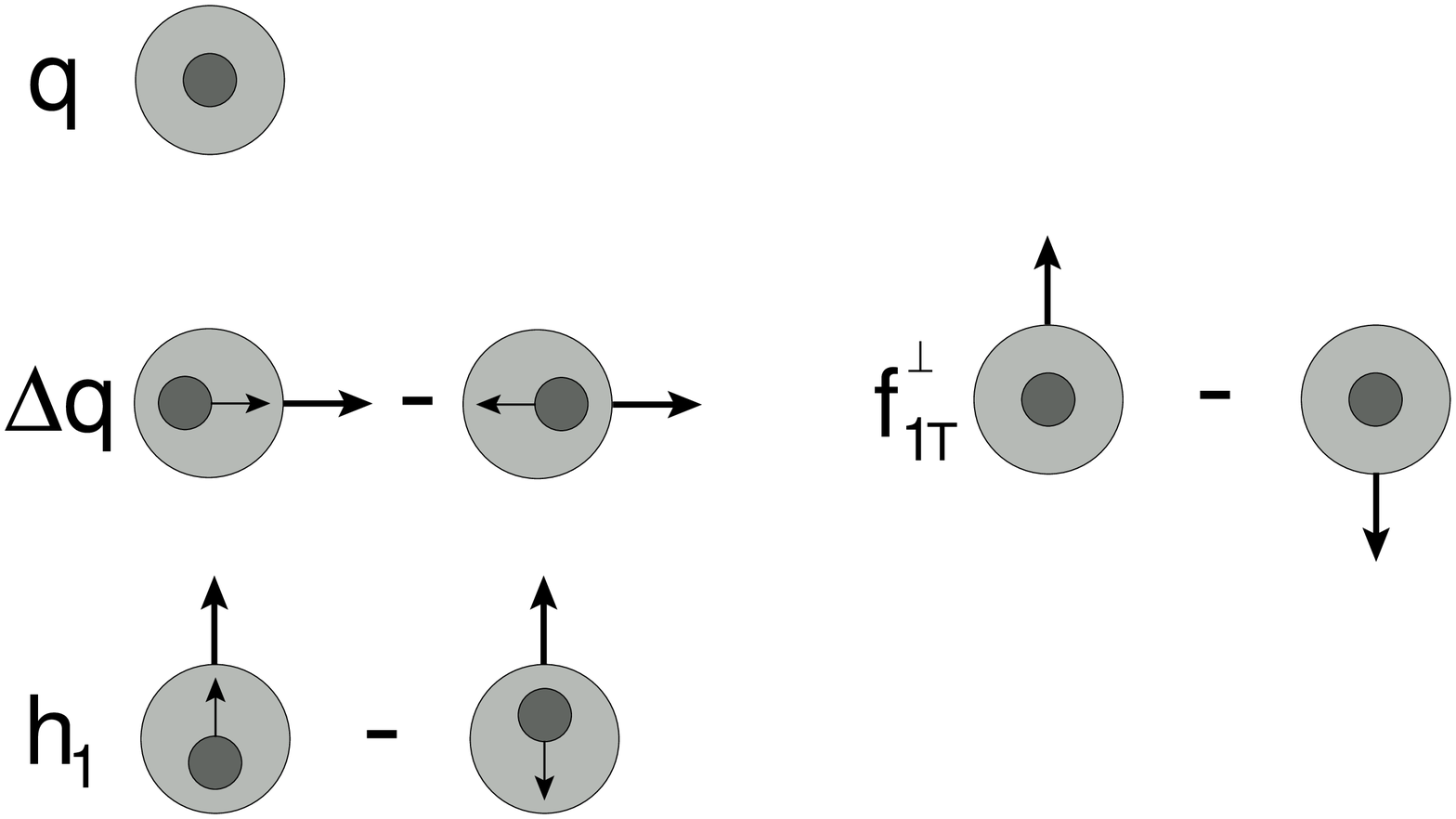,height=3.cm}}
    \put(8.3,0.){\psfig{figure=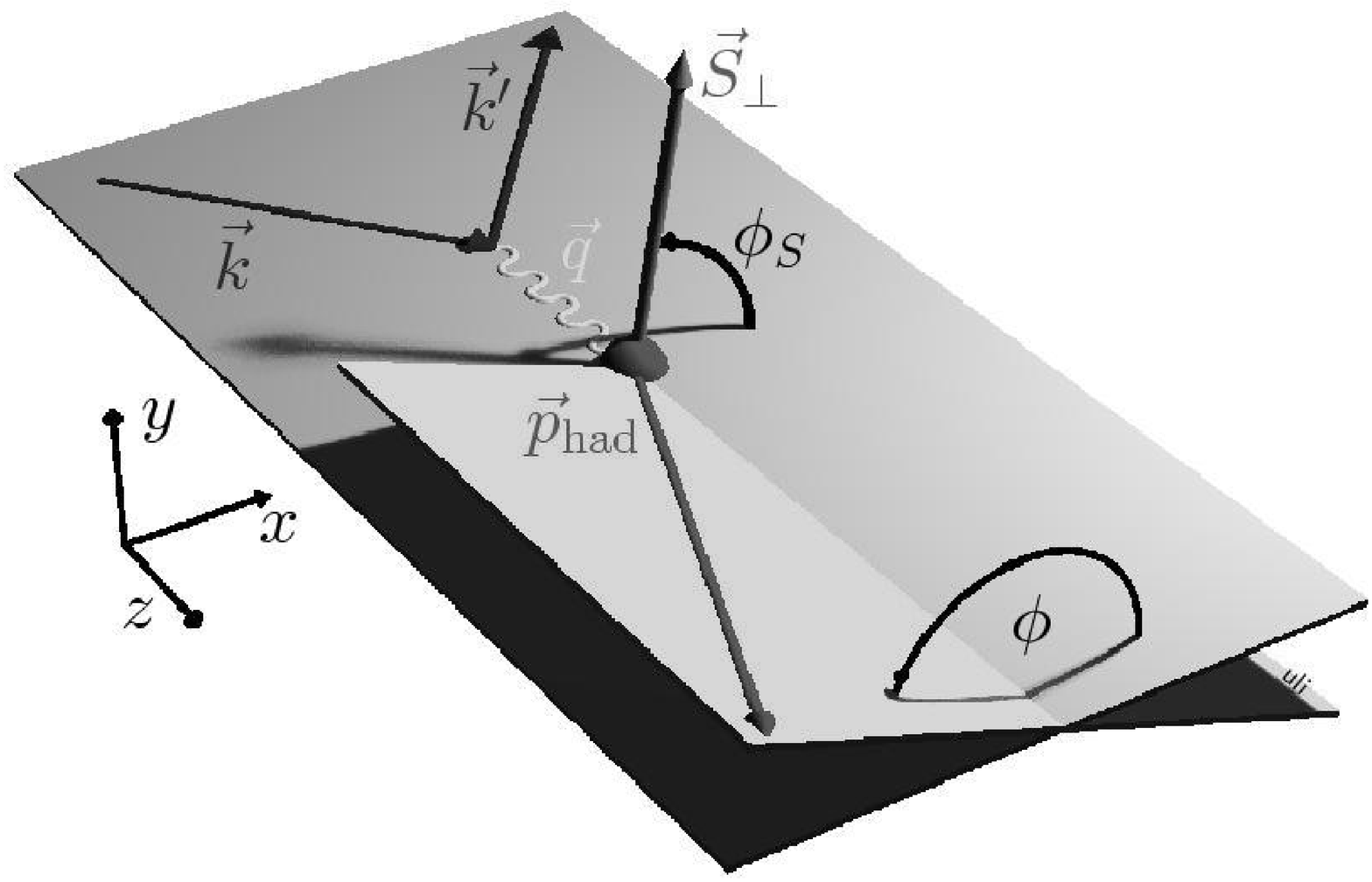,height=4.cm}}
    \put(3.5,-0.2){(a)}
    \put(12,-0.2){(b)}
  \end{picture}
  \caption{(a) Graphical illustration of the distribution functions.
    (b) Definition of the azimuthal angles.
    \label{fig:DFangles}}
\end{figure}

The four--momentum transfer to the target is a measure of the 
spatial resolution in the scattering process. DIS processes
have a momentum transfer larger than the mass of the nucleon and thus can 
resolve its constituents.
In the quark parton model in which the virtual photon is assumed to scatter 
incoherently off the quarks in the nucleon, the DIS cross section can be expanded
in terms of \textit{quark distribution 
functions}. In a frame in which the nucleon is moving towards the photon 
with ``infinite'' momentum the leading--twist distribution functions (DF) can be 
interpreted as probability densities dependent on the longitudinal quark 
momentum. Only three leading twist DFs survive the integration
over the intrinsic transverse quark momentum $p_T$.
The unpolarised DF $q(x)$ and the helicity DF 
$\Delta q(x)$ have already been explored for different quark flavours $q$ 
by several experiments \cite{Mar02} \cite{Air04}. (Here $x$ is the
dimensionless Bjorken scaling variable which can be identified with the
fractional momentum of the nucleon carried by the quark.)
The latter gives the probability to find a quark with its helicity parallel 
to the nucleon helicity.
The third DF is the chiral--odd \textit{transversity} \cite{Ral79} \cite{Art90}
\cite{Jaf92} function $h^q_1(x)$ which has no probabilistic interpretation in the
helicity basis but only in a basis of transverse spin eigenstates. In this basis
\linebreak 
it represents the degree to which the quarks are polarised along the proton's 
spin direction when the proton is polarised transversely to the virtual photon.
These three DFs are illustrated in Fig.~\ref{fig:DFangles}(a).
The light and dark grey circles represent the nucleon and the quark respectively
and the arrows indicate the spin directions. In each illustration the virtual 
photon is incident from the left side. The helicity DF and the transversity 
may differ, since the nucleon is a relativistic bound state and in the 
relativistic regime boosts and rotations do not commute.

In DIS chirality is conserved, thus transversity can be measured
only in a process in which it is combined with another chiral--odd object.
In semi--inclusive DIS produced hadrons are detected 
in addition to the scattered lepton, leading to the appearance in the cross 
section of fragmentation functions (FF) in conjunction with the DFs.
In unpolarised DIS for instance the cross section is proportional to the
product of the unpolarised DF $q(x)$ and the unpolarised FF
$D_1^q(z)$ which gives the probability density that a struck quark of flavour 
$q$ produces a certain final state hadron with the fractional energy $z$.
Using a transversely polarised nucleon target, the transversity enters the cross
section combined with the chiral--odd FF $H_1^{\perp q}(z)$ known as 
\textit{Collins} function \cite{Col93}. In addition, a second DF 
$f_{1T}^{\perp q}$ -- the so--called \textit{Sivers} function -- appears
in the cross section together with the unpolarised FF.
This DF relates the quark transverse momentum with the
transverse polarisation of the nucleon as depicted in Fig.~\ref{fig:DFangles}(a).
The property of the Sivers function to be odd under time reversal (T--odd) 
was believed to forbid its existence. But recently it was realised that 
final--state interactions via a soft gluon offer a mechanism to create the 
necessary interference of amplitudes \cite{Bro02} for the existence of the 
so--called ``na\"ive 
T--odd'' nature of the Sivers function, which means time--reversal without 
the interchange of initial and final states. An interesting explanation of 
a non--zero Sivers function is a non--vanishing orbital angular momentum of the 
quarks \cite{Bur02}.

\section{Azimuthal Asymmetries}

Since the Sivers and the Collins functions do not survive integration over
the intrinsic transverse momentum $p_T$ and the transverse momentum $k_T$
acquired in the fragmentation process, respectively, the tools to measure
the objects of interest are azimuthal asymmetries. These asymmetries depend on
two azimuthal angles $\phi$ and $\phi_S$ drawn in Fig.~\ref{fig:DFangles}(b).
The angle $\phi$ is defined between the lepton scattering plane containing the 
incoming and outgoing lepton and the hadron production plane spanned by the
produced hadron and the virtual photon. $\phi_S$ is the angle between the 
scattering plane and the transverse spin component of the target nucleon.

The luminosity normalised count rate asymmetry between opposite target spin 
states ($\up$,$\down$), weighting each event with the transverse momentum of the 
detected hadron $P_{h\perp}$, can be written as the sum of two sine functions,
as shown in Eq.~\ref{eq:asym}:
\begin{equation}\label{eq:asym}
  \frac{1}{S_\perp}
  \frac{\sum_{i=1}^{N^\up(\phi,\phi_S)} P_{h\perp\ i} -
    \sum_{i=1}^{N^\down(\phi,\phi_S)} P_{h\perp\ i}}
       {N^\up(\phi,\phi_S) + N^\down(\phi,\phi_S)} =
       A_\mathrm{UT}^{\sin(\phi+\phi_S)}\sin(\phi+\phi_S) +
       A_\mathrm{UT}^{\sin(\phi-\phi_S)}\sin(\phi-\phi_S)\ .
\end{equation}
The amplitudes of each sine term are proportional in leading order to the 
product of a DF and a FF: 
\begin{equation}
  A_\mathrm{UT}^{\sin(\phi+\phi_S)} \sim \sum_q e_q^2 \cdot
  h_1^q(x) \cdot H_1^{\perp(1)q}(z) 
  \quad\mbox{and}\quad
  A_\mathrm{UT}^{\sin(\phi-\phi_S)} \sim \sum_q e_q^2 \cdot
  f_{1T}^{\perp(1)q}(x) \cdot D_1^{q}(z)\ .
  \vspace*{-0.2\baselineskip}
\end{equation}
Here $S_\perp$ is the transverse polarisation of the target and the subscript UT 
indicates the unpolarised beam and the transversely polarised target. 
The superscript $(1)$ denotes the $p_T^2$-- or $k_T^2$--moment of the DF or FF, 
respectively. 
The $P_{h\perp}$ weighting is performed in order to avoid assumptions 
\cite{Kot97} about the quark transverse momentum dependencies. For unweighted 
asymmetries these assumptions are necessary to solve the convolution integral
\cite{Mul96} over $p_T$ and $k_T$ in which the product of DF and FF appears.
The sine--moments $A^{\sin(\phi\pm\phi_S)}$ of the asymmetry
were extracted performing a two--dimensional fit in order to minimise uncertainty
from systematic correlations.

\section{Requirements}

When measuring transverse spin asymmetries in semi--inclusive DIS three 
components are necessary. First of all a high energy lepton 
beam is needed which is provided by the HERA positron storage ring at DESY with 
an energy of 27.5 GeV. The positron beam interacts with the internal hydrogen gas
target of the HERMES experiment \cite{Ack98}. The hydrogen nuclei are 
transversely polarised
with an average polarisation (preliminary) of $0.74\pm0.06$ (systematic). 
The third required device is the
HERMES spectrometer which is used for the detection of the scattered leptons and 
the produced hadrons. This spectrometer provides lepton identification with an 
average efficiency of 98\% at a hadron contamination of less than 1\%.
Identification of certain hadron types is performed with
a Ring-Imaging \v Cerenkov detector (RICH) which allows
the efficient identification of charged pions, kaons, and protons over
almost the complete momentum range and hence leads to a very clean pion sample. 
The two photons of the decay of a neutral pion cause a pair of clusters
in the calorimeter. Both clusters have an energy larger than 1 GeV and 
cannot be assigned to a charged track in the detector. 
The invariant mass of the two clusters was required to be in an interval 
around the $\pi^0$ mass. The sidebands were used to evaluate the combinatorial 
background.

\section{Results}

\begin{figure}	
  \centerline{\psfig{figure=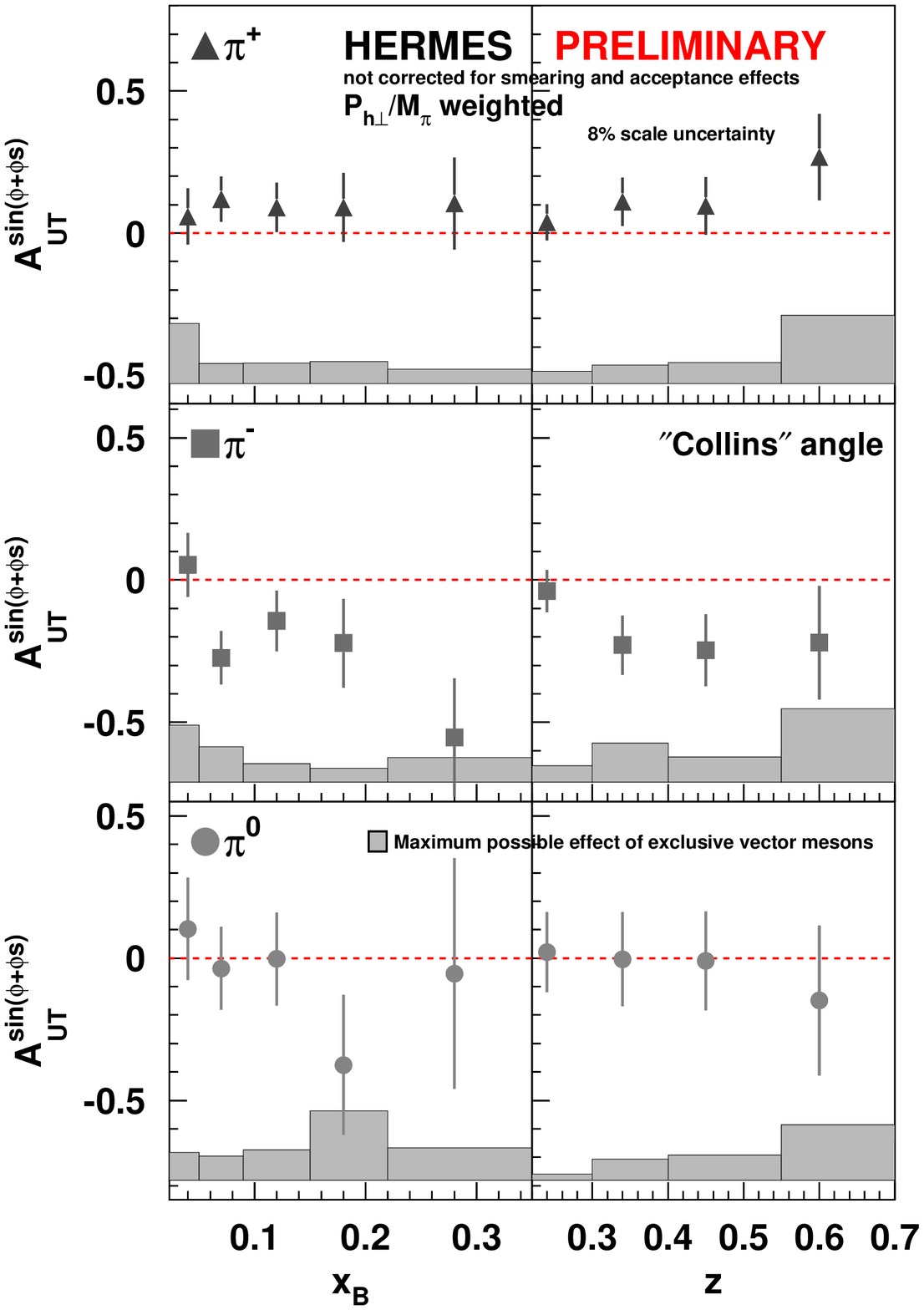,height=8cm}
    \hspace*{0.5cm}
    \psfig{figure=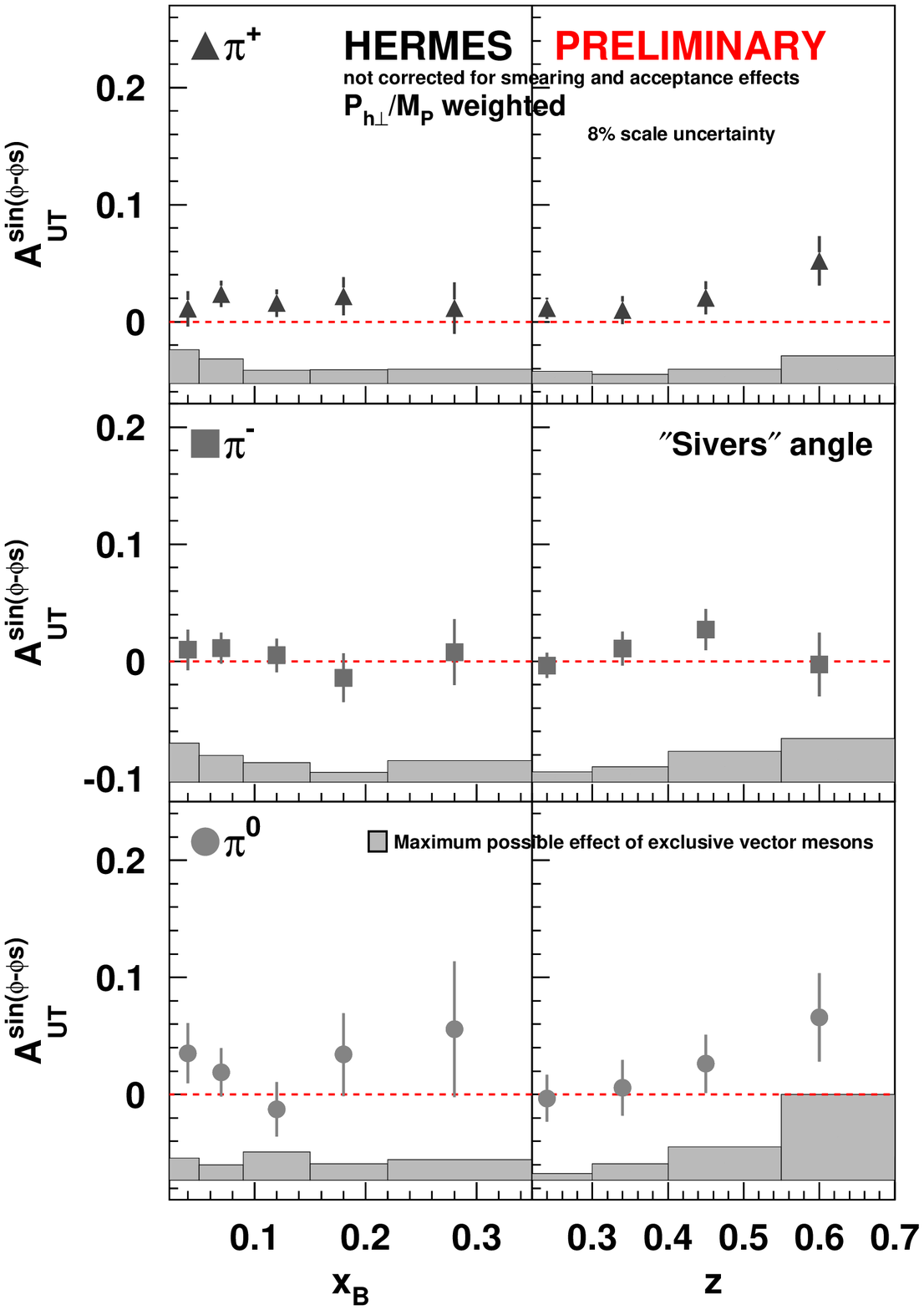,height=8cm}}
  \caption{Asymmetries $A_\mathrm{UT}^{\sin(\phi\pm\phi_S)}$ for
    $\pi^+$, $\pi^-$, and $\pi^0$ depending on the kinematic variables $x$ and 
    $z$.
    \label{fig:asyms}}
\end{figure}

In Fig.~\ref{fig:asyms} the measured asymmetries 
$A_\mathrm{UT}^{\sin(\phi\pm\phi_S)}$ are plotted for the production of
$\pi^+$, $\pi^-$, and $\pi^0$ mesons depending on $x$ and $z$.
The asymmetries $A_\mathrm{UT}^{\sin(\phi+\phi_S)}$ containing the product 
of transversity and Collins function
are positive for $\pi^+$, negative for $\pi^-$, and consistent with zero
for $\pi^0$. It is surprising that the magnitude of the $\pi^-$ 
asymmetry is at least as large as that for $\pi^+$, which was not the case
for the measurements on longitudinally polarised targets in the past 
\cite{Air03}. Positive asymmetries $A_\mathrm{UT}^{\sin(\phi-\phi_S)}$ are 
measured for $\pi^+$ and $\pi^0$ whereas the moment is consistent with zero 
for $\pi^-$. 
The grey error bands represent the maximum possible effect of
pions coming from the decay of diffractive vector mesons. At present there is 
little theoretical guidance on how to treat the contributions of the 
diffractive vector mesons. Thus an optional ``interpretive'' uncertainty is 
assigned which is based on a conservative estimate of 1 for the asymmetries 
of the vector mesons. These asymmetries cannot 
be measured with the existing data set due to acceptance limitations. 
Ongoing studies will reduce this uncertainty by roughly a factor of three
in the final analysis.

\section{Outlook}

The aim of the measurement of azimuthal asymmetries is the extraction of the
transversity and the Sivers function. 
The decoupling of the DF and FF using their different kinematic dependencies
is only possible with a larger data set. 
%
Another possibility is to use information about the FF. Parametrisations for 
the unpolarised FF $D_1^q(z)$
are sufficiently known for some types of produced hadrons \cite{Kre00}. An
extraction of the Sivers function is therefore already possible with the
existing asymmetry moments. Due to fundamental time reversal symmetry of
QCD the Sivers function is predicted to have the opposite sign in Drell--Yan 
compared to DIS \cite{Col02} -- a prediction which needs to be tested 
experimentally. 
The transversity extraction will be possible with the results of other 
experiments e.g., $e^+e^-$ annihilation experiments like BELLE and BABAR, 
will provide information about the Collins FF.

The HERMES experiment is continuing data taking with the transversely 
polarised target until summer 2005. Therefore a significant reduction of the
statistical uncertainties is expected. Also the COMPASS experiment at CERN
has accumulated data with a transversely polarised target. In the near future
new results will elucidate the properties of quarks in transversely polarised
nuclei.

\end{document}